\documentclass[aps,prb,twocolumn,showpacs]{revtex4}
\usepackage{graphicx}
\usepackage{hyperref}
\begin{document}
\title{Formation of clean dimers during gas-source growth of Si(001)}
\author{D.R.Bowler} \altaffiliation[Also at ]{London Centre for
Nanotechnology, Department of Physics and Astronomy, University
College London, Gower Street, London WC1E 6BT}
\email{david.bowler@ucl.ac.uk}
\homepage{http://www.cmmp.ucl.ac.uk/~drb/}
\affiliation{Department of Physics and Astronomy, University College London,
Gower Street, London WC1E 6BT, UK}

\date{\today}

\begin{abstract}
Elevated temperature STM measurements have shown that one key phase
during gas-source homoepitaxy of Si(001) is the formation of clean Si
ad-dimers from hydrogenated ad-dimers, though the mechanism for this
formation is unknown.  We present \textit{ab initio} density
functional calculations designed to explore this mechanism.  The
calculations show that there is a pathway consistent with the
experimentally observed reaction rates, which proceeds via a
meta-stable intermediate, and is effectively irreversible.  This
result fills a vital gap in our understanding of the atomic-scale
details of gas-source growth of Si(001).
\end{abstract}

\maketitle

\section{\label{sec:intro}Introduction}

Gas-source growth of Si(001) using hydrogen-based precursors (such as
SiH$_4$, silane, and Si$_2$H$_6$, disilane) is of great scientific and
technological interest\cite{Bron1993,Wang1994,Owen1997a,Owen1997b} ---
in particular, hydrogen can act as an effective surfactant, and has
been shown to reduce roughness and intermixing during growth of Ge/Si
alloys and pure Ge on Si(001)\cite{Ohtani1994}.  Understanding the
reactions that occur and the intermediate structures that are formed
during this growth will enable greater control of surface and
interfaces during growth.  STM observations of the growth of Si(001)
from disilane, both at room temperature following
anneals\cite{Bron1993,Wang1994} and at elevated
temperature\cite{Owen1997a,Owen1997b}, along with careful electronic
structure calculations\cite{Owen1997a,Owen1997b,Bowler1996} have
mapped out the growth pathway.  A key observation in this pathway is
that the islands formed during growth are \textit{clean}, while the
substrate remains covered with a certain amount of
hydrogen\cite{Owen1997a}.  The fundamental building block in
gas-source growth is the clean ad-dimer (as opposed to solid-source
growth, where fast-moving ad-atoms are key\cite{Swartz1996}); yet, the
mechanism to form such clean dimers from the hydrogenated dimers that
occur naturally during gas-source growth is unknown. In particular,
they are observed to form at 450K while desorption from the
monohydride phase occurs at 790K, indicating that their formation must
be completely different to the desorption of hydrogen from the
monohydride phase.  In this paper, we present a first-principles
investigation of the mechanism for formation of clean ad-dimers from
hydrogenated dimers, with the aim of explaining how these form at a
comparatively low temperature.

\begin{figure*}
\includegraphics[width=0.3\textwidth,clip]{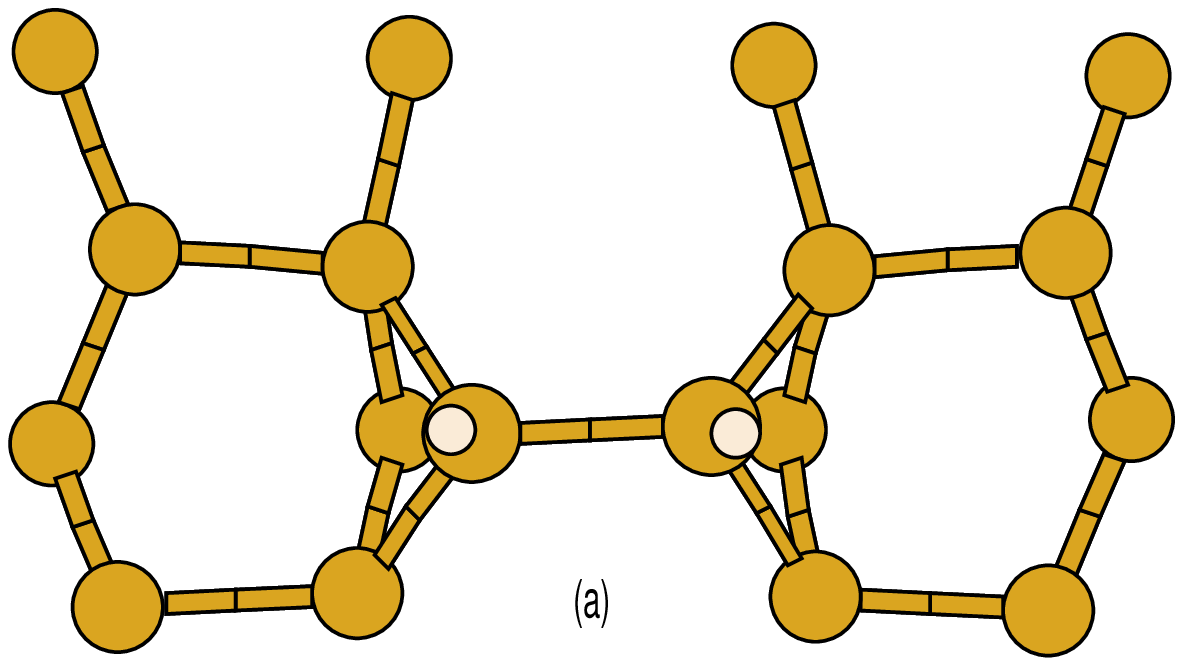}
\hskip0.025\textwidth
\includegraphics[width=0.3\textwidth,clip]{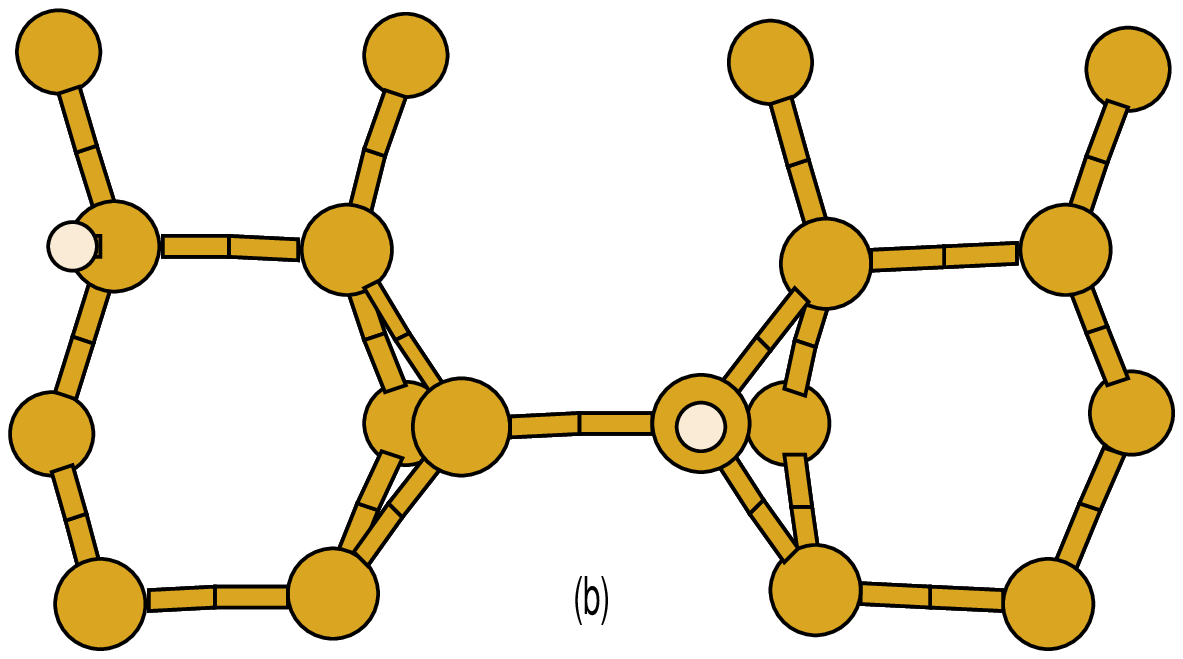}
\hskip0.025\textwidth
\includegraphics[width=0.3\textwidth,clip]{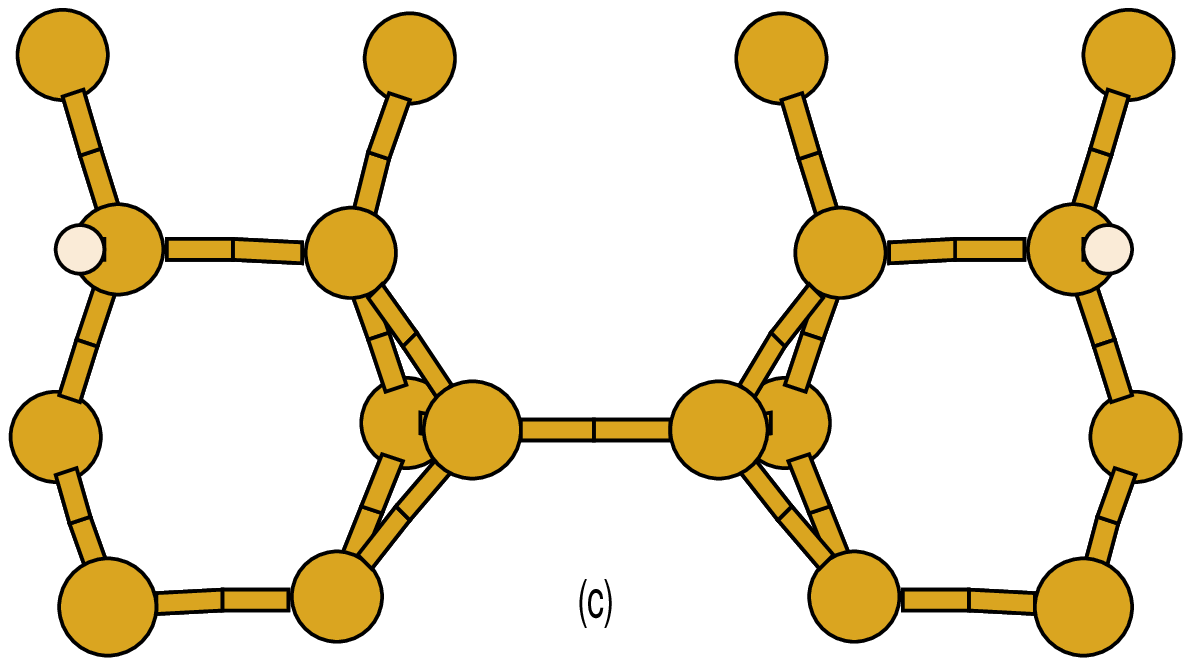}
\caption{\label{fig:EndsStruc}Structures of: (a) the starting point, with
a hydrogenated ad-dimer; (b) the end point for the first diffusion event
with one hydrogen on the substrate; and (c) the final point with a clean
ad-dimer and both hydrogens on the substrate.}
\end{figure*}

Disilane (which is used in preference to silane as it decomposes more
easily) adsorbs on Si(001) as SiH$_3$ (which soon breaks down to form
SiH$_2$) or SiH$_2$\cite{Bron1993}, sometimes with accompanying
hydrogen.  These SiH$_2$ groups\cite{Bowler1996} start to diffuse at
400--500K\cite{Owen1997a}.  When two groups are on adjacent dimer
rows, they react to form a hydrogenated ad-dimer (that is, an ad-dimer
with both dangling bonds saturated with hydrogen, illustrated in
Fig.~\ref{fig:EndsStruc}~(a)) over the trench between the dimer
rows\cite{Owen1997a}.  This then decomposes to form clean ad-dimers
and hydrogen on the surface at around
450K\cite{Bron1993,Wang1994,Owen1997a}, via a pathway to be
investigated in this paper.  A hydrogenated ad-dimer (which is the
starting point) is illustrated in Fig.~\ref{fig:EndsStruc}~(a), along
with a partially hydrogenated ad-dimer (the result of the first part
of the pathway) in Fig.~\ref{fig:EndsStruc} (b) and a clean ad-dimer
(the final point) in Fig.~\ref{fig:EndsStruc} (c).  Once formed, the
clean ad-dimers diffuse\cite{Goringe1997} and form a square feature,
which is believed to be the precursor to dimer
strings\cite{Owen1997c}, followed by short strings of
dimers\cite{Wang1994} which later increase to form larger
islands\cite{Wang1994,Owen1997b}.

The calculations to be presented are based on density functional
theory (DFT) in the generalized gradient approximation (GGA), with a
plane wave basis set and pseudopotentials.  We have searched for
possible pathways both by applying constraints to specific atoms (for
instance constraining a hydrogen to lie in a given plane) and by using
the nudged elastic band technique
(NEB)\cite{Jonsson1998,Henkelman2000b} which allows accurate
determination of reaction barriers given an initial approximation to a
pathway.  One key result is that the dehydrogenation proceeds via a
meta-stable intermediate state (this is discussed fully in
Section~\ref{sec:meta} and illustrated in Fig.~\ref{fig:MetaStruc}).

The rest of the paper is organised as follows: the next section gives
details of the computational techniques used; this is followed by a
detailed discussion of the structure of the meta-stable state which
plays a key role in the dehydrogenation; the diffusion pathways are
then presented, looking at the mechanism for both hydrogens, followed
by a conclusion section.

\section{\label{sec:details}Computational Details}

The theory underlying DFT\cite{Hohenberg1964,Kohn1965} and their
application to electronic structure calculations have been extensively
reviewed\cite{Jones1989}, as has the use of pseudopotential and
plane-wave techniques\cite{Payne1992}.  The calculations in this paper
were performed using the VASP code\cite{Kresse1996}, using the
standard ultra-soft pseudopotentials\cite{Vanderbilt1990} that form
part of the code.  The approximation we use for exchange-correlation
energy is the generalised-gradient approximation (GGA) due to Perdew
and Wang (PW91)~\cite{Wang1991,Perdew1992}.  We chose the GGA rather
than the local density approximation (LDA) rather deliberately.  As
the barriers that we will be calculating are sensitive to bonding and
stretched bonds, and the GGA is known to be rather more accurate in
these situations (LDA generally overbinds), we considered its use to
be essential for this work.

We use periodic boundary conditions, as is standard for DFT
calculations with plane-waves, and we therefore used a periodic slab
for the surface, with a vacuum layer between the slabs.  Our
simulations were performed within a unit cell two dimers long and two
dimer rows wide, with five layers of Si (the bottom of which was
terminated in hydrogen and constrained to remain fixed in bulk-like
positions).  The vacuum gap of 6.9~\AA\ is equivalent to five atomic
layers of Si, and provides sufficient isolation between vertical
periodic images.  We used a plane wave cutoff of 150~eV and a $2
\times 2 \times 1$ Monkhorst-Pack k-point mesh.  All these parameters
were tested, and found to converge energy differences to better than
0.01~eV.  The system contains an even number of electrons, but has
various saddle points which might involve unpaired electrons, so we
checked the effect of performing spin-polarised calculations for these
points. The effect was found to be negligible (both on energies and
geometries) and so was not used in the calculations.

To investigate the diffusion pathways, we used two techniques: first,
constraining the diffusing hydrogen to lie in a particular plane, and
calculating static energies for different locations of the hydrogen;
second, the Nudged Elastic Band
method\cite{Jonsson1998,Henkelman2000b}.  This second method requires
the simultaneous relaxation of a number of images of the system, which
can be done in parallel.  However, this has the potential to become
extremely computationally intensive, which is why we chose to use the
smallest realistic unit cell (with four dimers in the surface).

The initial exploration of the system used the first method (static
calculations, constraining the hydrogen).  It was using this method
that the meta-stable state (discussed in Sec.~\ref{sec:meta}) was
found, and it is unlikely that it would have been found using the NEB
without significant effort (for instance performing simulated
annealing on the initial images of the system) or using a more
complicated technique such as the dimer method\cite{Henkelman1999}.
The diffusion barriers presented in the paper were all calculated
using a variant of the original technique which actively seeks the
saddle point, the climbing image NEB\cite{Henkelman2000}, with 8
images relaxed in the chain.

While we have calculated the diffusion barriers, we have not
calculated attempt frequencies, which have been assumed to be $10^{13}
\ \mathrm{sec}^{-1}$, typical for diffusion processes.  While DFT-GGA
is sufficiently accurate to calculate reaction barriers to within
0.1~eV, it is not able to predict attempt frequencies
accurately\cite{Blochl1990}; for instance, in previous work on
solid-source growth of Si(001), it was shown that a factor of five in
the attempt frequency was required to understand the results, but was
not accurately predicted\cite{Smith1997}.

\section{\label{sec:meta}The Meta-stable Intermediate}

\begin{figure}
\includegraphics[width=0.9\columnwidth,clip]{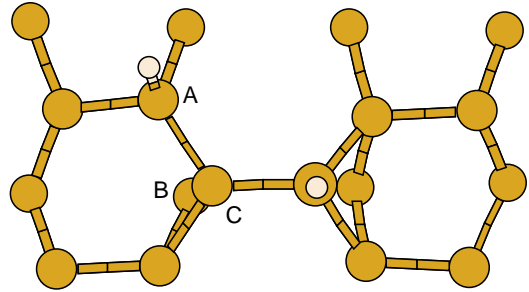}
\caption{\label{fig:MetaStruc}The structure of the meta-stable state
which provides a low energy pathway for dehydrogenation.  The
diffusing hydrogen is bonded to a substrate dimer (A), which has
broken one bond to a second-layer silicon (B).  The ad-dimer (C) is
now partly clean, and has formed a bond to the second-layer silicon
(B) left by the substrate dimer.}
\end{figure}

The lowest energy diffusion pathway, and the only one which has an
energy barrier which is in line with the temperature at which the
dehydrogenation is observed to occur, proceeds via a meta-stable
intermediate.  This is an unusual and rather important structure, and
will be discussed in detail in this section.  It is illustrated in
Fig.~\ref{fig:MetaStruc}, and should be contrasted with the
hydrogenated ad-dimer illustrated in Fig.~\ref{fig:EndsStruc}~(a).

The atom labelled `A' in Fig.~\ref{fig:MetaStruc} is one of the
substrate dimers, to which the hydrogenated (and clean) ad-dimer is
bonded; `B' is a second layer atom in the substrate, to which the
substrate dimer is normally bonded (for instance in
Fig.~\ref{fig:EndsStruc}~(a)); and `C' is the ad-dimer atom itself,
which is now clean (having started hydrogenated).  The hydrogen is now
bonded to `A', which has broken its bond to `B', while `C' has formed
a bond to `B' (not easily seen, owing to the geometry).

In terms of the bonding of the atoms, the atoms A, B and C are all
saturated as they are in Fig.~\ref{fig:EndsStruc}~(a) --- the bonding
has merely cycled around (so that the A--B bond is now an A--H bond
and a B--C bond, while the C--H bond is now a C--B bond).  It is this
saturation that gives the structure its stability.  While some of the
bond angles are rather strained (in particular, the bonds associated
with B and C make $60^\circ$ angles) the bond lengths are all close to
equilibrium, and there are no further broken bonds, leading to an
energy difference of 0.57~eV relative to the starting point, but no
more.  It is interesting to note that there are other structures where
$60^\circ$ bond angles are found during growth of Si(001), which also
exhibit stability which might seem counter-intuitive\cite{Bowler1996}.

In terms of the formation of this structure, as we shall see in the
next section, there is not a large barrier.  The H never has to move a
long distance from either A or C, leading to relatively strong bonds
being present at all times; the second-layer atom B moves up slightly;
and while the substrate dimer atom A and the ad-dimer atom C do move
up and down respectively, they do this gradually while maintaining
their bonding.  It is this relatively small perturbation on the
overall structure, and the ease with which it is reached, which allows
the formation of this state, and gives it its importance.

\section{\label{sec:paths}Diffusion Pathways}

In this section, we describe the diffusion pathways that we have
explored with DFT calculations.  For simplicity, and because it is
likely to be physically realistic, we allow the hydrogens on the ends
of the dimer to diffuse off independently --- i.e. we consider the
diffusion off one end of the ad-dimer while the other hydrogen remains
on the ad-dimer.  Then we allow the remaining hydrogen to diffuse off
the now partially-hydrogenated ad-dimer onto the substrate.  In order
to avoid the complications of spin and half-filled bands, we maintain
both hydrogens in the unit cell at all times (the first hydrogen to
diffuse off stays on the substrate, illustrated in
Fig.~\ref{fig:EndsStruc}(b)).  The three stable points of the process
(fully hydrogenated ad-dimer, partially hydrogenated ad-dimer with a
hydrogen on the substrate and clean ad-dimer with both hydrogens on
the substrate) are illustrated in Fig.~\ref{fig:EndsStruc}.  The
atomic positions during the diffusion pathways are presented below in
an aggregated form (due to space constraints): only the position of
atoms which move significantly are shown.  All of the atomic
structures at each step in all the diffusion pathways are available
elsewhere\cite{DeHydroPage}.

The experimental data that we are comparing against comes from two
separate experiments: first, where the Si(001) surface was exposed to
a dose of disilane (Si$_2$H$_6$), annealed at different temperatures
for different times, and then observed at room temperature in
STM\cite{Bron1993, Wang1994}; second, where an elevated-temperature
STM was used to observe the results of dosing with disilane in real
time at different temperatures\cite{Owen1997a,Owen1997b}.  The results
of both these types of experiment are identical: around 450K, clean,
non-rotated dimers are formed over the trench between dimer rows.  In
other words, the monohydride dimers lose their hydrogen to the
substrate in a matter of minutes at this temperature (for instance, an
anneal to 470K for two minutes led to the dehydrogenation of all
ad-dimers\cite{Wang1994}).  Assuming an attempt frequency of $10^{13}\
\mathrm{sec}^{-1}$ and a successful dehydrogenation rate of 1/60~Hz,
we obtain a barrier of 1.28~eV.  This changes by about 0.03~eV if the
rate is doubled or halved, giving us a good estimate of the likely
reaction barrier.

\subsection{\label{sec:first}The First Hydrogen}

\begin{figure}
\includegraphics[width=0.9\columnwidth,clip]{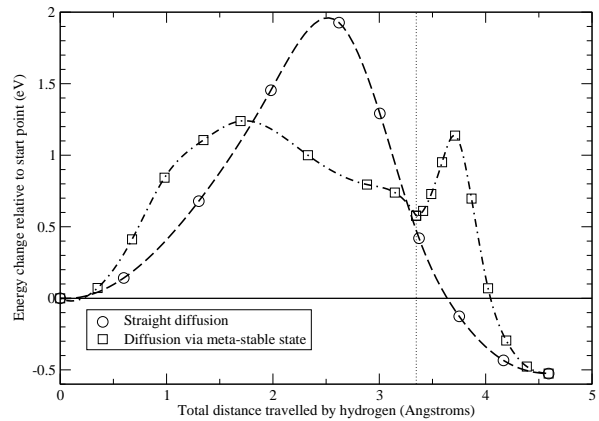}
\caption{\label{fig:FirstHDiff}A graph showing energy barriers for the
diffusion for the first hydrogen off the ad-dimer.  The energies are
given relative to the starting point, while the \textit{x}-axis gives
the distance from the starting point of the hydrogen.  The open
circles show direct diffusion (proceeding without the meta-stable
intermediate state).  The open squares show the diffusion via the
meta-stable state (whose position is marked with a vertical dotted
line at 3.35\AA, and whose structure is shown in
Fig.~\protect\ref{fig:MetaStruc}).  The lines (long dashes for direct
diffusion and dash-dotted for diffusion via the meta-stable state) are
spline fits to the data, and are given as guides to the eye.}
\end{figure}

There are two diffusion paths considered for the first hydrogen
diffusing off the ad-dimer: a direct diffusion path; and diffusion via
the meta-stable considered in Sec.~\ref{sec:meta} and shown in
Fig.~\ref{fig:MetaStruc}. We will discuss these separately, starting
with the direct diffusion, and then contrast their results.

\begin{figure}
\includegraphics[width=0.9\columnwidth,clip]{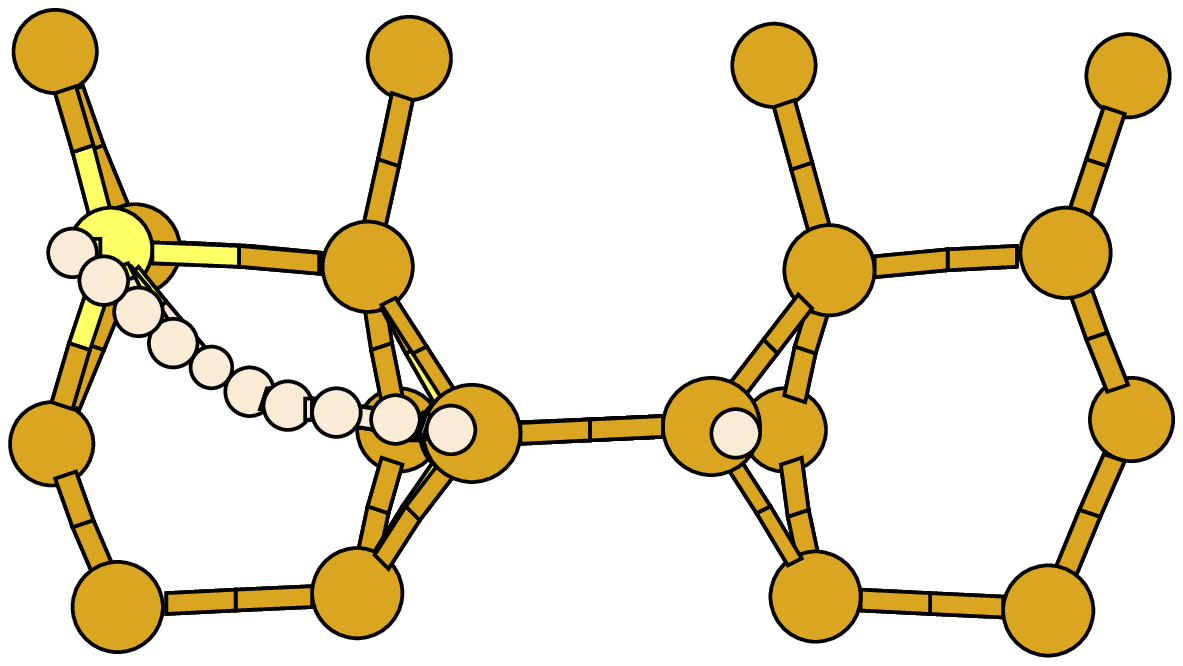}
\includegraphics[width=0.9\columnwidth,clip]{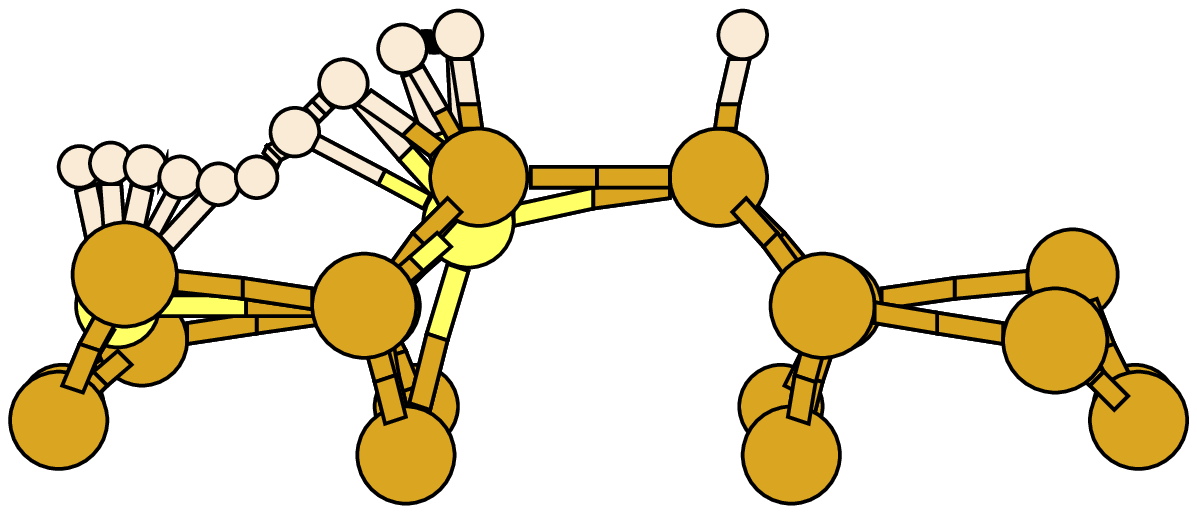}
\caption{\label{fig:H1S2ESaddle}The path of the first hydrogen in the
direct diffusion path from above (top) and the side (bottom).  All the
hydrogen positions are shown, along with the initial and final
positions of the atoms to which the hydrogen bonds (with the final
position shown in a lighter shade).  The final position of the
ad-dimer cannot be seen in the top image since it is directly below
the initial position.  Bonds (or lack of bonds) are produced by the
imaging software, and should not be taken as definite indications.}
\end{figure}

The diffusion barrier for direct diffusion is shown in
Figure~\ref{fig:FirstHDiff}, with open circles and long dashes.  The
barrier is 1.93~eV, which is extremely high; the reason for this can
be seen from the atomic positions, which are illustrated in
Figure~\ref{fig:H1S2ESaddle}.  At the saddle point, the ad-dimer bond
is extended greatly (from 2.51~\AA\ at the start to 2.82~\AA) while
the bond from the hydrogen to the ad-dimer is more extended (from
1.51~\AA\ at the start to 1.85~\AA).  Inspecting the charge density,
it is clear that the ad-dimer remains bonded (though weakly) and that
the H has made a weak bond to the substrate dimer (which is 2.32~\AA\
away) as well as maintaining a slightly weakened bond to the ad-dimer.
It is this lengthening and weakening of bonds at the saddle point that
causes the high barrier.  Assuming Arrhenius behaviour and an attempt
frequency of $10^{13}$~Hz, we find a hopping rate of $\sim10^{-10}
\mathrm{sec}^{-1}$ at 450K, which is many orders of magnitude below
the observed rate.

The diffusion barrier into and out of the meta-stable state is also
shown in Fig.~\ref{fig:FirstHDiff}, with open squares and a dot-dashed
line.  The barrier from the start to the meta-stable state is 1.24~eV,
while the barrier from the meta-stable state to the end is 0.56~eV
(and the reverse path, from the meta-stable state to the start is
0.66~eV).

\begin{figure}
\includegraphics[width=0.9\columnwidth,clip]{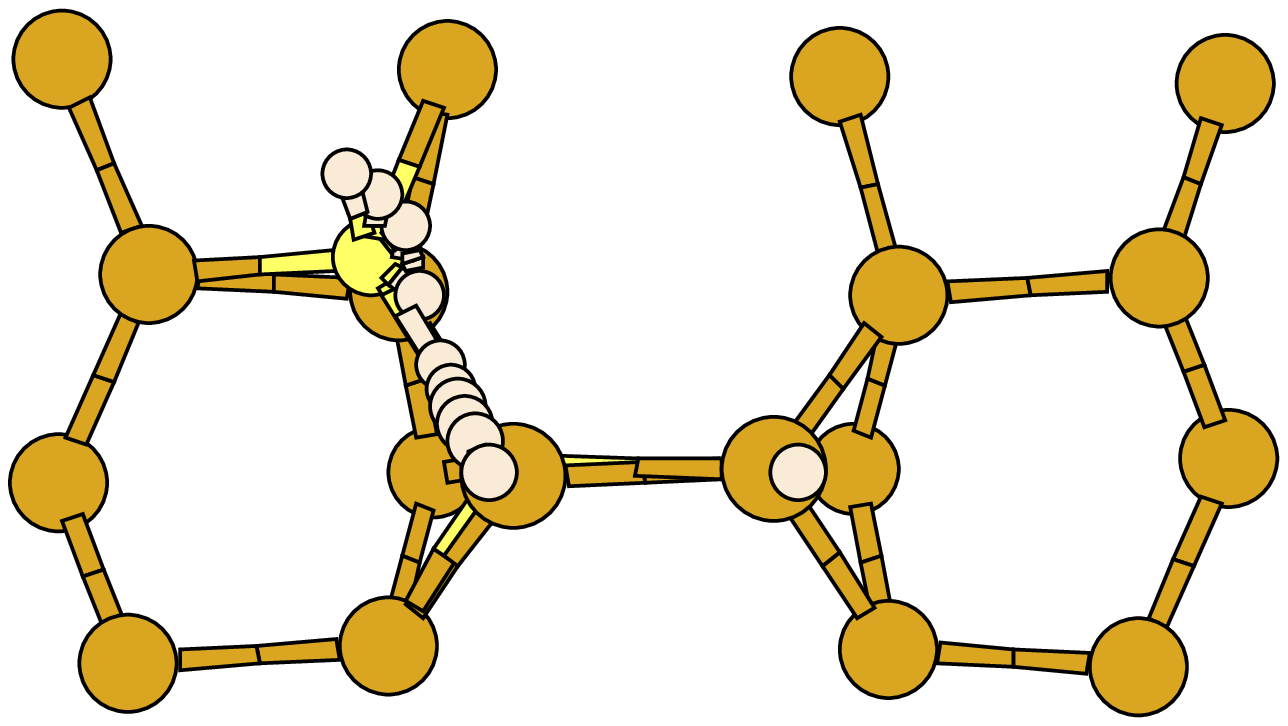}
\includegraphics[width=0.9\columnwidth,clip]{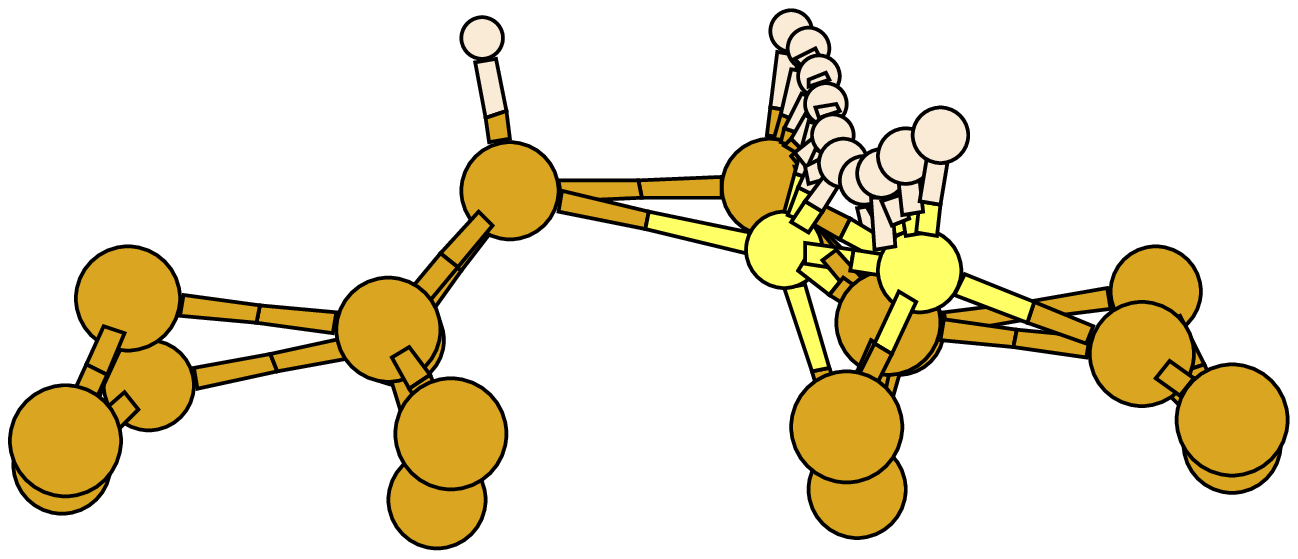}
\caption{\label{fig:S2MSaddle}The path of the first hydrogen from the
ad-dimer to the meta-stable state shown in views from above (top) and
the side (bottom).  All the hydrogen positions are shown, along with
the initial and final positions of the atoms to which the hydrogen
bonds (with the final position shown in a lighter shade).  The side
view is shown rotated by 180$^\circ$ relative to the view from above
as the image is clearer.  The final position of the ad-dimer cannot be
seen in the top image since it is directly below the initial position.
Bonds (or lack of bonds) are produced by the imaging software, and
should not be taken as definite indications.}
\end{figure}

\begin{figure}
\includegraphics[width=0.9\columnwidth,clip]{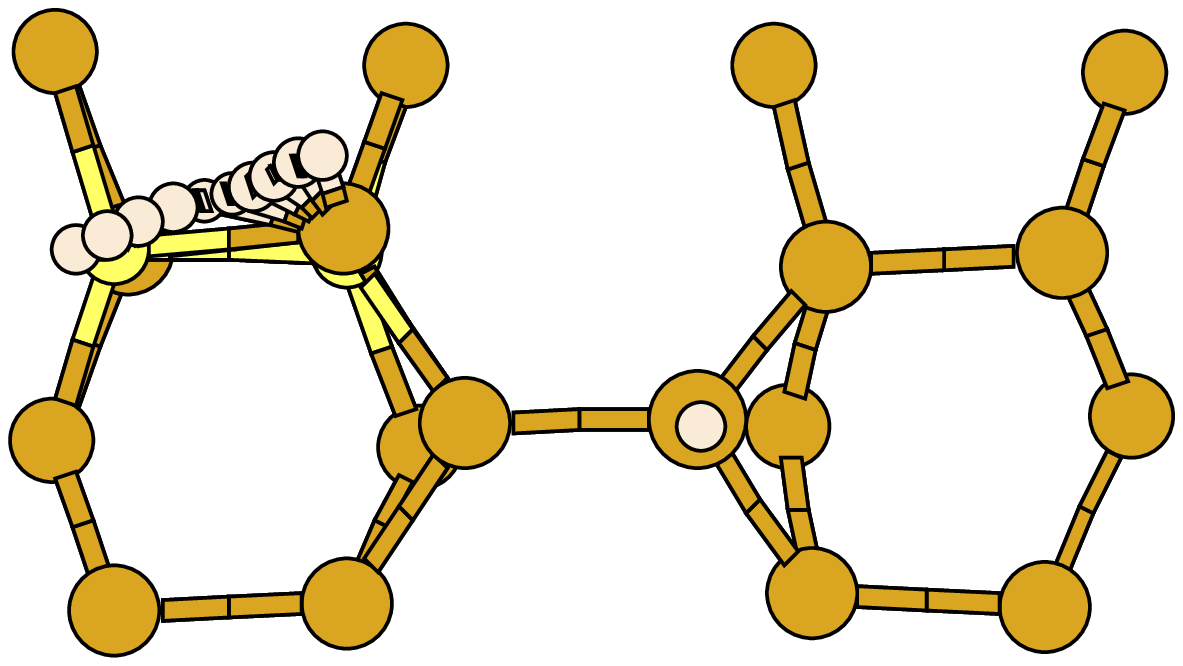}
\includegraphics[width=0.9\columnwidth,clip]{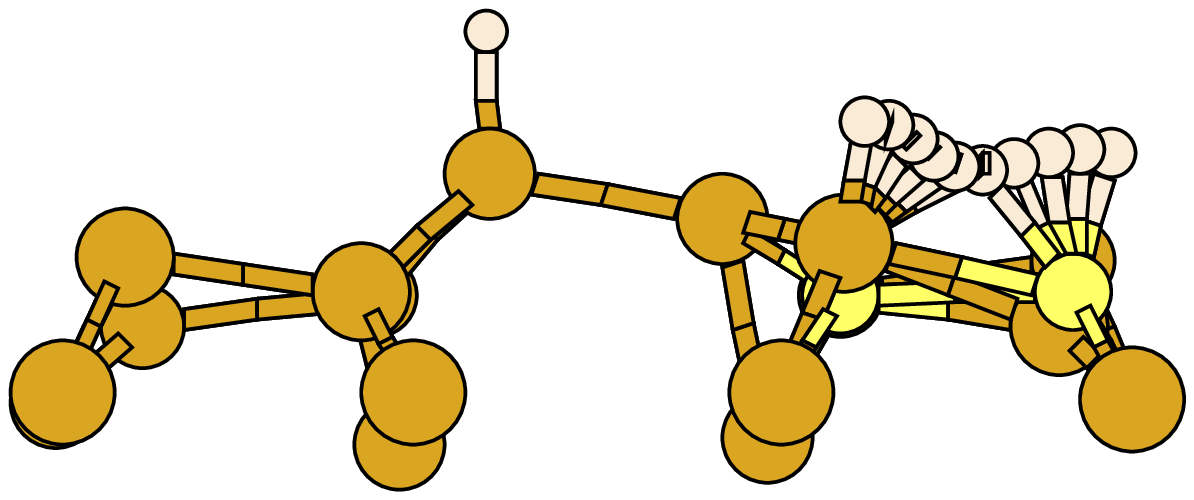}
\caption{\label{fig:M2ESaddle}The path of the first hydrogen from the
meta-stable state to the substrate dimer shown in views from above
(top) and the side (bottom).  All the hydrogen positions are shown,
along with the initial and final positions of the atoms to which the
hydrogen bonds (with the final position shown in a lighter shade).
The side view is shown rotated by 180$^\circ$ relative to the view
from above as the image is clearer.  Bonds (or lack of bonds) are
produced by the imaging software, and should not be taken as definite
indications.}
\end{figure}

The pathway from the starting position to the meta-stable state (shown
in Fig.~\ref{fig:S2MSaddle}) involves considerable rearrangement:
first, the hydrogen inserts into the bond between the ad-dimer
(labelled `C' in Fig.~\ref{fig:MetaStruc}) and the substrate dimer
(labelled `A' in Fig.~\ref{fig:MetaStruc}); at the saddle point, the
hydrogen is 1.65~\AA\ from the ad-dimer, and 1.94~\AA\ from the
substrate dimer (compared to an equilibrium distance of 1.51~\AA),
while the distance between the ad-dimer and the substrate dimer is
3.10~\AA\ (compared to 2.48~\AA\ at the start); second, after the
hydrogen has transferred to the substrate dimer, the ad-dimer bonds to
a \textit{second layer} atom in the substrate (labelled `B' in
Fig.~\ref{fig:MetaStruc}); third, the substrate dimer bonds back to
the ad-dimer and breaks its bond to the second layer atom in the
substrate.  The first part of this rearrangement is the area where
most of the energy change happens: the energy actually falls by about
0.6~eV during the second and third parts of the rearrangement.

The pathway from the meta-stable state to the end position (shown in
Fig.~\ref{fig:M2ESaddle}) is much simpler, involving only the movement
of the hydrogen from one end of the substrate dimer to the other,
while the substrate dimer `C' reforms its bond to the second layer
atom `B'.  At the saddle point, the hydrogen is 1.72~\AA\ from the
substrate atom and 2.05~\AA\ from the end atom.

The barrier of 1.24~eV from the starting point to the meta-stable
state fits extremely well with the observed temperature behaviour: at
450K with an attempt frequency of $10^{13}\ \mathrm{sec}^{-1}$, it
would correspond to a hopping rate of 0.044~Hz, or one hop every 23
seconds.  But this is only \textit{into} the meta-stable state, and
there are \textit{two} low energy paths out of that.  The hopping rate
from the meta-stable state to the end state is $\sim 4\times
10^{6}$~Hz, while from the meta-stable state to the start state is
$\sim 2\times 10^{5}$~Hz, so that only 10\% of meta-stable states
would return to the starting point.  We also expect that the
equilibrium populations of the start and end states would differ,
since the end state is 0.53~eV lower than the start (roughly, at 450K,
we would expect a relative population about $10^{6}$ times higher in
the lower state).  There is also the question of whether the hydrogen
could return, via the meta-stable state, from the end to the start.
The barrier from the end point \textit{back} to the meta-stable state
it is 1.66~eV, making it extremely unlikely that the hydrogen would
return to the meta-stable state (and even if it did, it would be ten
times more likely to drop back to the end state than to return to the
start state).  Clearly, it is the low barrier from the starting state
to the meta-stable state that allows the first part of the
dehydrogenation of the ad-dimer to proceed, and the energy difference
between the start and end points, as well as the high barrier out of
the end state that makes the reaction effectively irreversible.

\subsection{\label{sec:second}The Second Hydrogen}

Once the first hydrogen has diffused off the ad-dimer, we retain it on
the substrate, as shown in Figure~\ref{fig:EndsStruc}(b).  This is
computationally convenient (as it maintains a filled set of bands) but
also physically reasonable: hydrogen does not begin diffusing along
the dimer rows on Si(001) at an appreciable rate until about 550K with
a barrier of 1.68~eV\cite{Owen1996}.

\begin{figure}
\includegraphics[width=0.9\columnwidth,clip]{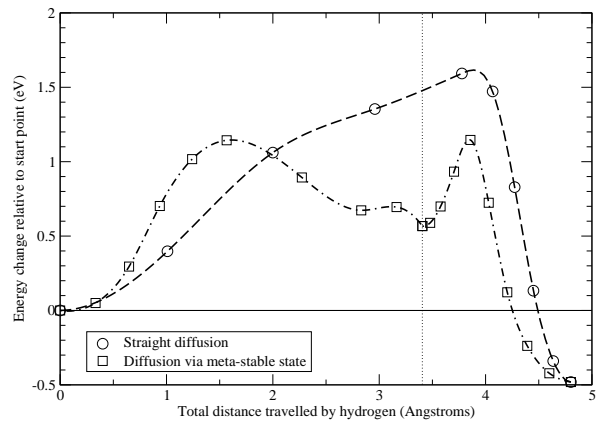}
\caption{\label{fig:SecondHDiff}A graph showing energy barriers for
the diffusion of the second hydrogen off the ad-dimer (with the first
hydrogen already on the substrate).  The energies are given relative
to the starting point, while the \textit{x}-axis gives the distance
from the starting point of the hydrogen.  The open circles show direct
diffusion (proceeding without the meta-stable intermediate state).
The open squares show the diffusion via the meta-stable state (whose
position is marked with a vertical dotted line at 3.41\AA, and whose
structure is shown in Fig.~\protect\ref{fig:MetaStruc}).  The lines
(long dashes for direct diffusion and dash-dotted for diffusion via
the meta-stable state) are spline fits to the data, and are given as
guides to the eye.}
\end{figure}

As with the first hydrogen, the second hydrogen can diffuse either
directly, or via a meta-stable state, which is exactly equivalent to
the meta-stable state for the first hydrogen (shown in
Fig.~\ref{fig:MetaStruc}) and therefore not illustrated here.  As
before, we will discuss these results separately, starting with the
direct diffusion.

\begin{figure}
\includegraphics[width=0.9\columnwidth,clip]{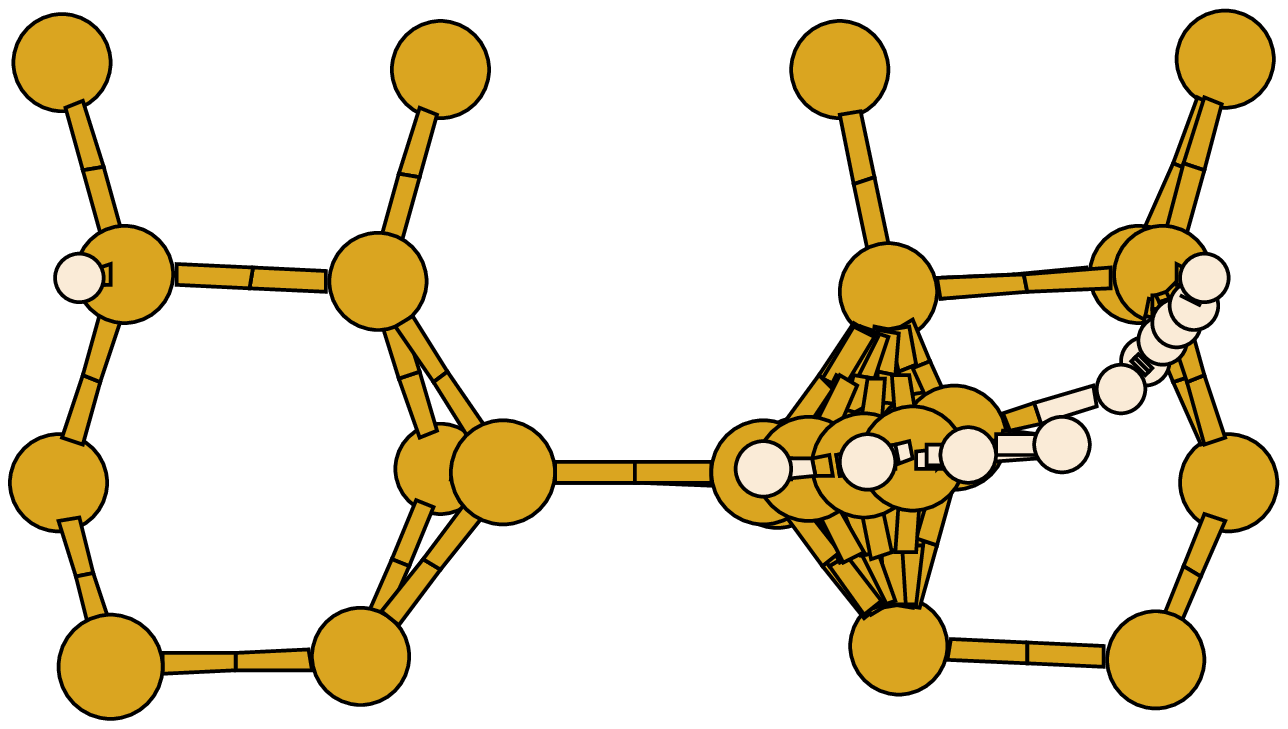}
\includegraphics[width=0.9\columnwidth,clip]{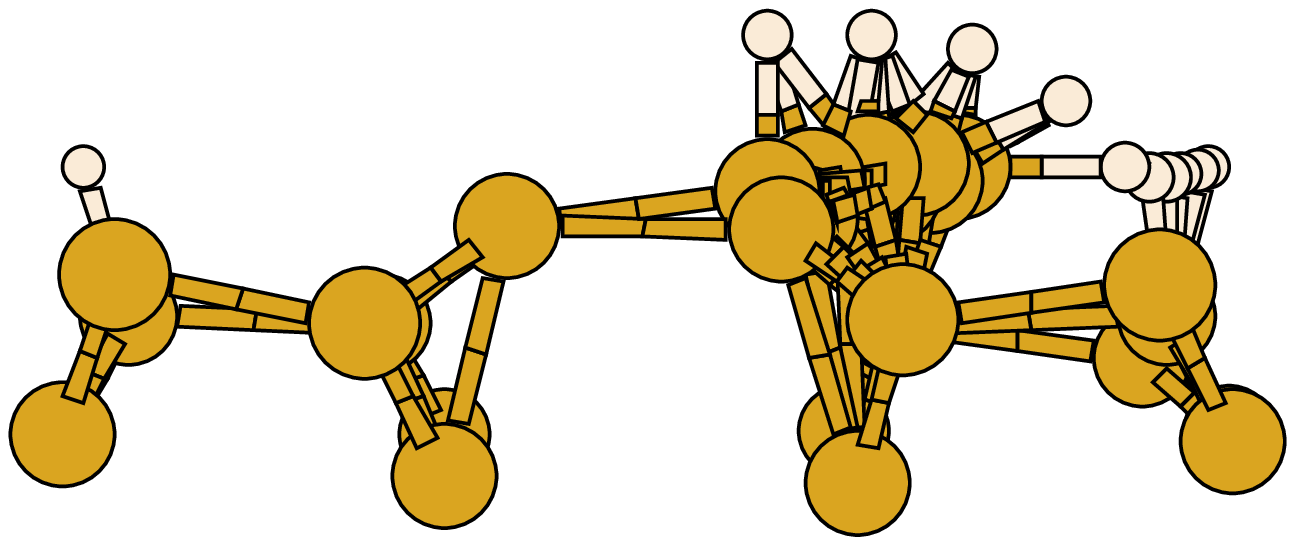}
\caption{\label{fig:H2S2ESaddle}The path of the second hydrogen in the
direct diffusion path shown in views from above (top) and the side
(bottom).  All the hydrogen positions are shown, as are the positions
of the atoms to which the hydrogen bonds.  Bonds (or lack of bonds)
are produced by the imaging software, and should not be taken as
definite indications.}
\end{figure}

The diffusion barrier for direct diffusion is shown in
Figure~\ref{fig:SecondHDiff}, plotted with open circles and dashes.
The shape is much broader than for the first hydrogen's direct path,
with a lower barrier of 1.59~eV.  The reason for this can be seen in
part in the atomic positions, which are shown in
Figure~\ref{fig:H2S2ESaddle}.  This is a little more confusing than
previous plots, as the positions both of the hydrogen and the ad-dimer
atom to which it is bonded have been plotted.  As the hydrogen moves
across towards the substrate dimer (the end point), the bond between
the silicon atoms in the ad-dimer breaks, with the atom that the
hydrogen is bonded to following the hydrogen as it diffuses.  At the
saddle point, the hydrogen is 1.67~\AA\ from the ad-dimer and
2.03~\AA\ from the substrate dimer, while the distance between
ad-dimer atoms is 4.48~\AA.  Beyond this point, the hydrogen transfers
to the substrate, and the ad-dimer reforms slowly.

The bond in the clean or partially clean ad-dimer is not as strong as
the other bonds to the substrate, which explains why the energy cost
for breaking it is relatively small, and why this pathway is followed
in contrast to the pathway for the first hydrogen.  Even with the
reduced barrier, the hopping rate at 450K will be $\sim
4\times10^{-6}\ \mathrm{sec}^{-1}$, which is still far too low to be
consistent with the experimental observations.

The diffusion barrier into and out of the meta-stable state is also
shown in Figure~\ref{fig:SecondHDiff}, plotted with open squares and
dot-dashed lines.  The barrier from the start to the meta-stable state
is 1.14~eV, while the barrier from the meta-stable state to the end is
0.58~eV (and the reverse path, from the meta-stable state to the start
is also 0.58~eV).  The atomic positions are almost identical to those
for the first diffusion (shown in Figs.~\ref{fig:S2MSaddle} and
\ref{fig:M2ESaddle}) and are not shown (though these figures, and many
other pieces of supplementary information such as animations of the
processes can be found elsewhere\cite{DeHydroPage}).

The barrier of 1.14~eV is 0.1~eV lower than the barrier for the first
hydrogen, suggesting that once the first hydrogen has diffused off the
ad-dimer, the second will follow slightly more quickly; it is still in
excellent agreement with observed experimental behaviour.  The
barriers from the meta-stable state to the start and end states are
now identical, meaning that 50\% of meta-stable states will return to
the starting state.  However, the end state is 0.48~eV lower in energy
than the start, so that (as before with the first hydrogen) we would
expect the population in thermal equilibrium at 450K to be about
$10^{6}$ times higher in the end state than the start state.  The
barrier from the end state back to the meta-stable state is 1.63~eV,
which again makes the reaction effectively irreversible.  Of course,
the clean ad-dimer can also diffuse away along the trench between
dimer rows, with a barrier of 1.15~eV\cite{Goringe1997}, which would
make the reforming of the hydrogenated ad-dimer impossible.

\section{\label{sec:conc}Conclusions}

We have presented \textit{ab initio} calculations, modelling the
diffusion of hydrogen off a hydrogenated ad-dimer, which is a key
stage in gas-source growth of Si(001).  We have shown that the
diffusion proceeds via a meta-stable intermediate, and that the energy
barriers calculated (1.24~eV for the first hydrogen and 1.14~eV for
the second hydrogen) are in excellent agreement with temperatures at
which these features are observed in experiment.

We have used the climbing image nudged elastic band method to find the
diffusion barriers, and have found it to be extremely effective,
particularly for the direct diffusion which was difficult to model
simply by picking a single constraint.  However, the problem of
exploring phase space is still a difficult one, as the existence of
the meta-stable state (which was discovered through application of a
single constraint) shows.  There are techniques for exploring energy
surfaces, such as the dimer method\cite{Henkelman1999} and variants on
hyperdynamics\cite{Voter1997}, but there is still a large amount of
work to be done in this field.

\begin{acknowledgments}
We gratefully acknowledge useful discussions with Dr J.H.G.Owen,
Dr. C.M.Goringe and Prof. M.J.Gillan.  DRB thanks the Royal Society
for funding through a University Research Fellowship.  Calculations
were performed at the HiPerSPACE Centre, UCL which is partly funded by
JREI grant JR98UCGI.  We also acknowledge the Jonsson group in
University of Washington for making their implementation of the NEB
and climbing image NEB within VASP publicly available.
\end{acknowledgments}

\bibliography{DeHydro}

\end{document}